\documentclass[twocolumn,preprintnumbers,
superscriptaddress,amsmath,floatfix]{revtex4}

\usepackage{amssymb}  
\usepackage{graphicx,subfigure}
\usepackage{exscale}
\usepackage{textcomp}
\usepackage{enumerate}
\usepackage{amsmath}
\usepackage{amssymb}
\usepackage{color,graphicx}
\usepackage[dvipsnames]{xcolor}
\usepackage{bbm}
\usepackage[pdfstartview=XYZ,
bookmarks=true,
colorlinks=true,
linkcolor=blue,
urlcolor=blue,
citecolor=blue,
pdftex,
bookmarks=true,
linktocpage=true, 
hyperindex=true
]{hyperref}
\usepackage{orcidlink}

\usepackage{color}

\usepackage[normalem]{ulem}  

\newcommand{\nsat}{n_0}

\newcommand\si{\sigma}

\newcommand\om{\omega}

\newcommand{\be}{\begin{equation}}
\newcommand{\ee}{\end{equation}}
\newcommand{\bea}{\begin{eqnarray}}
\newcommand{\eea}{\end{eqnarray}}
\newcommand{\ba}[1]{\begin{array}{#1}}
\newcommand{\ea}{\end{array}}

\newcommand{\Msolar}{\ensuremath{{\rm M}_\odot}}

\newcommand{\sliver}{\kern 0.07em} 

\newcommand{\chiEFT}{\raisebox{0.2ex}{\ensuremath{\chi}}\text{EFT}}

\newcommand{\fm}{\text{fm}}

\newcommand{\ptrans}{P_\mathrm{tr}}

\begin{document}

\title{Nuclear and Hybrid Equations of State in Light of the Low-Mass Compact Star in HESS J1731-347}

 \author{L. Brodie\,\orcidlink{0000-0001-7708-2073
}}
 \email{b.liam@wustl.edu}
 \affiliation{Department of Physics, Washington University in St.~Louis, St.~Louis, MO 63130, USA}
 
 \author{A. Haber\,\orcidlink{0000-0002-5511-9565}}
 \email{ahaber@physics.wustl.edu}
 \affiliation{Department of Physics, Washington University in St.~Louis, St.~Louis, MO 63130, USA}

\date{11 August 2023}

\begin{abstract}
We sample over $500$ relativistic mean-field theories constrained by chiral effective field theory and properties of isospin-symmetric nuclear matter and test them against known stellar structure constraints. This includes a recent mass and radius measurement of a compact object in supernova remnant HESS J1731-347, with an unusually low mass of $M=0.77^{+0.20}_{-0.17}\,\Msolar$ and a compact radius of $R=10.4^{+0.86}_{-0.78}$ km. We show that none of the sampled nuclear models meet all constraints at the $68\,\%$ credibility level, but that hybrid equations of state with a quark matter inner core and nuclear outer core easily can. This indicates a tension between astrophysical constraints and low-energy nuclear theory.

\end{abstract}

\maketitle

\section{Introduction}
\label{sec:intro}
Constraining the equation of state (EOS) of dense matter and the phase diagram of quantum chromodynamics (QCD) is one of the great tasks of modern theoretical physics \cite{Bogdanov:2019owz,Maggiore:2019uih,Evans:2021gyd,Lovato:2022vgq, Bogdanov:2022faf}. By combining astrophysical observations with theoretical calculations, we can study the otherwise inaccessible high-density and low-temperature part of the QCD phase diagram. Each conjectured EOS must be thoroughly tested against current astrophysical constraints \cite{Riley:2021pdl,Pang:2021jta,Miller:2021qha,superlight} to validate its high-density behavior \cite{Nandi:2018ami,Essick:2020flb,Greif:2020pju,Raaijmakers:2021uju,Legred:2021hdx,Alford:2022bpp,Salinas:2023nci}. First principle calculations, e.g., chiral effective field theory (\chiEFT) at low densities \cite{Lynn:2015jua,Tews:2018kmu,Lonardoni:2019ypg, Drischler:2020yad} and perturbative QCD at high densities \cite{Somasundaram:2022ztm,Gorda:2022lsk,Annala:2019puf,Annala:2017llu}, put additional constraints on the EOS.
Astrophysical observations of stellar masses, radii, and tidal deformabilities (which strongly depend on the compactness of a star, i.e.,~the ratio of mass and radius) play a major role in constraining the EOS, because they can be computed directly from the EOS by solving the Tolman-Oppenheimer-Volkoff equations from general relativity. 

In the last decade, compact stars with masses greater than two solar masses that were measured accurately via Shapiro delay \cite{Antoniadis:2013pzd,Cromartie:2019kug,Fonseca:2021wxt} have been used to rule out many models. More recently, measurements of stellar radii \cite{Miller:2019cac,Riley:2019yda,Miller:2021qha,Riley:2021pdl} have started to restrict the allowed mass-radius parameter space.
In 2022, the central compact object (CCO) in HESS J1731-347 was reported to have an unexpectedly low mass of $M=0.77^{+0.20}_{-0.17}\,\Msolar$ and surprisingly compact radius of $R=10.4^{+0.86}_{-0.78}\,$km \cite{superlight}.
In this publication, we test how well models of nuclear matter describe known stellar structure constraints, including the CCO in HESS J1731-347.

For this purpose, we sample hundreds of relativistic mean-field theories (RMFTs) constrained at low densities by \chiEFT\ following the procedure we developed in Refs.~\cite{Alford:2022bpp,Alford:2023rgp}. RMFTs are  useful tools that allow us to  
model nuclear matter across the wide range of densities and temperatures found in neutron stars and their mergers \cite{Walecka1974,Boguta:1977xi,Serot:1984ey,Glendenning1996}. 
In addition to the EOS, RMFTs provide a consistent framework for chemical equilibration~\cite{Alford:2021ogv, Most:2022yhe}, response to magnetic fields~\cite{Broderick:2000pe,Haber:2014ula}, 
and  calculations of transport properties such as bulk viscosity~\cite{Alford:2019qtm,Alford:2019kdw}.

After investigating traditional nuclear models, we show that hybrid stars with an inner core of quark matter and an outer core of nuclear matter can meet all astrophysical constraints at the $68\,\%$ credibility level. To do so we combine a soft nuclear EOS, where the pressure does not rise rapidly with density leading to small stellar radii predictions, at low densities with a constant speed of sound model (CSS) for quark matter at high densities. We constrain the free parameters of the model using astrophysical constraints and show the compatibility of our results with other recent studies, for example a Bayesian analysis of the speed of sound in compact stars \cite{Ecker:2022xxj} or constraints from perturbative QCD \cite{Komoltsev:2021jzg,Gorda:2022jvk}. Our approach differs from the proposed solution of a strange quark star in Refs.~\cite{superlight,DiClemente:2022wqp,Oikonomou:2023otn,Horvath:2023uwl} because we study hybrid EOSs that model a nuclear matter phase below a critical transition density instead of postulating pure strange quark matter stars. A similar idea was presented in Ref.~\cite{Tsaloukidis:2022rus}, where the authors focus on hybrid stars that form so-called ``twin stars" \cite{Glendenning:1998ag,Schertler:2000xq,Li:2022ivt,Li:2023zty}, which achieve a small radius by branching off a stiffer nuclear EOS. Twin stars might be disfavored according to Ref.~\cite{Legred:2021hdx}. 

Our work motivates the further improvement of low-energy nuclear theory predictions and astrophysical observations. It also highlights new avenues to directly constrain low-energy nuclear effective field theories using astrophysical observations. Furthermore, we show how the unconfirmed assumption of pure nucleonic compact stars can lead us to falsely exclude models of dense matter. 

This paper is organized as follows: In Sec.~\ref{sec:nuclear}, we investigate how the sampled RMFTs obey a wide range of astrophysical constraints, including the mass and radius measurement of the CCO in HESS J1731-347 presented in Ref.~\cite{superlight}. In Sec.~\ref{sec:hybrid}, we augment the RMFT with a phase transition to quark matter and show that these hybrid models are capable of reproducing all astrophysical measurements at the $68\,\%$ credibility level, while the nuclear models are only compatible at the $95\,\%$ credibility level. We end by presenting our conclusions in Sec.~\ref{sec:conclusions}. In all our calculations we use natural units, \mbox{$\hbar=c=k_B=1$}.

\section{Nuclear Equations of State}
\label{sec:nuclear}
\begin{figure}
    \centering
    \includegraphics[width=1\hsize]{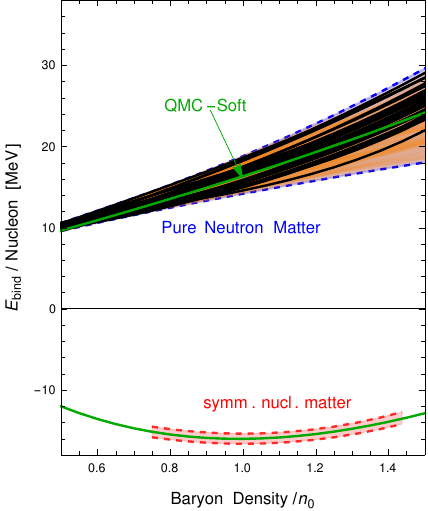}
    \caption{
       Binding energy per nucleon as a function of baryon number density in pure neutron matter. The blue \chiEFT\ uncertainty band is from Ref.~\cite{Tews:2018kmu}. We sample over $500$ RMFTs (orange lines) in this uncertainty band and compute the corresponding mass-radius curves. The $25$ samples that predict a maximum mass greater than two solar masses and go through all $95\,\%$ credibility contours of the astrophysical constraints described at the beginning of Sec.~\ref{sec:nuclear} are shown in black. The QMC-Soft EOS for pure neutron matter (upper panel) and symmetric nuclear matter (lower panel) is shown in green.}
    \label{fig:chipt_sample}
\end{figure}

In this section, we test how well nuclear EOSs can meet the following astrophysical constraints: 
\begin{itemize}
    \item The mass measurement of pulsar J0740+6620 from Ref.~\cite{Riley:2021pdl}: $M=(2.072\pm0.066)\,\Msolar$.
    \item The multimessenger constraints from Ref.~\cite{Pang:2021jta} using the NICER+XMM-Newton result Ref.~\cite{Miller:2021qha}, which combines NICER and XMM-Newton observations of pulsars, tidal deformability constraints from two gravitational-wave detections -- GW170817 and GW190425, and detailed modeling of the kilonova AT2017gfo and the gamma-ray burst GRB170817A.
    \item The mass-radius measurement of the CCO in supernova remnant HESS J1731-347 reported in Ref.~\cite{superlight} with a low mass of \mbox{$M=0.77^{+0.20}_{-0.17}\,\Msolar$} and compact radius of \mbox{$R=10.4^{+0.86}_{-0.78}\,$km}.
\end{itemize} 
We sample over $500$ RMFTs within the \chiEFT\ uncertainty band for pure neutron matter from Ref.~\cite{Tews:2018kmu} between baryon density $0.5\,\nsat$ and $1.5\,\nsat$, where $\nsat=0.16\,\mathrm{fm}^{-3}$ is the saturation density of isospin-symmetric nuclear matter. This guarantees that our RMFTs do not violate the best available theoretical constraints for pure neutron matter at zero temperature (Fig.~\ref{fig:chipt_sample}). 

The \chiEFT\ uncertainty band that we use can in principle widen further by taking different many-body interactions and computational methods into account \cite{Drischler:2017wtt,Piarulli:2019pfq,Lovato:2022apd}. These calculations predict uncertainty bands for the binding energy per nucleon  that are greater in magnitude and slope (note that the slope is proportional to the pressure) and therefore an even stiffer EOS. To allow for stars with smaller radii (i.e.,~low central pressures), we will perform our analysis with the \chiEFT\ uncertainty band of Ref.~\cite{Tews:2018kmu}. While pure neutron matter is currently best described by \chiEFT, the properties of isospin-symmetric nuclear matter are better constrained by other theories and experiments. We thus ensure that all sampled RMFTs reproduce known experimental and inferred properties of isospin-symmetric nuclear matter around saturation density \cite{Bethe:1971xm,Shlomo:2006incompressibility,Li:2019xxz}; see Ref.~\cite{Alford:2022bpp} for more details. 

The sampled RMFTs have the same interaction terms as the well-established IU-FSU RMFT \cite{Fattoyev:2010mx}. To test for model-induced biases we add additional meson self-interaction terms and re-fit the model to the pressure curve obtained without the additional interaction terms. We find variations smaller than the uncertainties introduced by \chiEFT\ which lead to only minor  changes in the mass-radius curve. 

\begin{figure}
    \centering
    \includegraphics[width=1\hsize]{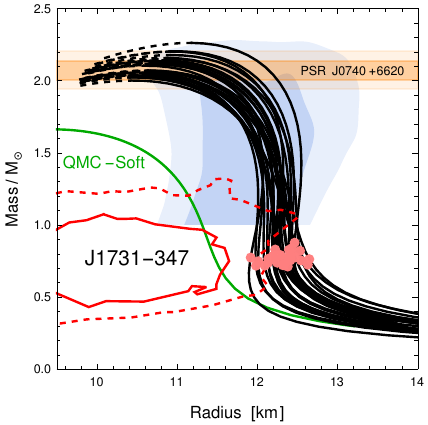}
    \caption{
       Mass-radius curves for the RMFTs sampled in Fig.~\ref{fig:chipt_sample} that obey all relevant astrophysical constraints at the $95\,\%$ credibility level. The orange-shaded bars show the $68\,\%$ credibility (dark shading) and $95\,\%$ credibility (light shading) mass measurement of pulsar J0740+6620 from Ref.~\cite{Riley:2021pdl}. 
    The blue-shaded area shows the 68\,\% credibility (dark shading) and 95\,\% credibility (light shading) multimessenger constraints from Ref.~\cite{Pang:2021jta} using the NICER+XMM-Newton result of Miller et al.~\cite{Miller:2021qha}. The red solid and dashed lines show the mass-radius contours for the CCO in HESS J1731-347 from Ref.~\cite{superlight} at $68\,\%$ and $95\,\%$ credibility, respectively. The pink dots indicate a central density of $n_B=2\,\nsat$, the uppermost density to which $\chiEFT$ might be reliable. The QMC-Soft EOS is shown in green.}
    \label{fig:chipt_sample_MR}
\end{figure}

To predict the mass and radius of compact stars, we solve the Tolman-Oppenheimer-Volkoff (TOV) equation. This requires the knowledge of the EOS in chemical or so-called beta-equilibrium. We use the RMFT framework to compute the EOS in chemical equilibrium, which does not require any additional fixing of parameters. Below a baryon density of $n_B\approx0.25\,\nsat$ we attach the GPPVA(TM1e) crustal EOS from CompOSE \cite{compose:gppva_tm1e}. This crustal EOS combines the Baym-Pethick-Sutherland EOS \cite{Baym:1971pw} for densities below $n_B=0.002\,\mathrm{fm}^{-3}$ with a Thomas-Fermi calculation \cite{Grill:2014aea} using the TM1e RMFT \cite{Shen:2020sec} for the inner crust. The complete sampling, crust attachment procedure, and Lagrangian can be found in Ref.~\cite{Alford:2022bpp}. Varying the crust transition density between $0.1\,\nsat$ and $0.5\,\nsat$ and using different crustal EOSs influences $R(1.4\,\Msolar)$ and $R(0.7\,\Msolar)$ on the order of $\Delta R\approx 100\,$m and has no major influence on the conclusions presented in this paper.
Out of more than 500 RMFTs we sample, none of them obey all astrophysical constraints at the $68\,\%$ credibility level, and only $25$ can fulfill the $95\,\%$ credibility constraints (Fig.~\ref{fig:chipt_sample_MR}). Fig.~\ref{fig:chipt_sample} depicts the uncertainty band from \chiEFT\ \cite{Tews:2018kmu} (bounded by blue, dashed lines) of the binding energy of pure neutron matter, which can not be directly probed experimentally. While the orange lines show the entirety of our sampled models, the black solid lines represent the $25$ models that are able to obey all astrophysical constraints described at the beginning of this section at the $95\,\%$ credibility level. For an EOS to predict a star with a small radius of $10-11\,$km and mass of $M\approx0.77\,\Msolar$, the EOS must have small pressures at low densities. This implies a relatively flat curve in the \chiEFT\ uncertainty band since the pressure can be obtained as a derivative of the binding energy with respect to the density. To meet the two-solar-mass constraint, the binding-energy curve must be as flat as possible and then rapidly increase, so the pressure becomes large enough to sustain a two-solar-mass star. Even the most extreme RMFTs that we sampled were not able to achieve this and thereby fulfill all constraints at the $68\,\%$ credibility level.
For most of our models, it is only the irregular shape of the mass-radius contour reported in Ref.~\cite{superlight} that allows
them to meet all constraints at the $95\,\%$ credibility level.
We show this in Fig.~\ref{fig:chipt_sample_MR}, where we plot the mass-radius curves of the $25$ ``successful" models from our sampling procedure. The solid and dashed red lines and the blue and orange contours depict the astrophysical constraints described at the beginning of this section.

Should further measurements confirm the central value reported in Ref.~\cite{superlight}, there will be a notable tension with low-energy nuclear physics computed from first principles via \chiEFT, especially given that several $\chiEFT$ higher order calculations imply an even stiffer EOS at low densities \cite{Drischler:2017wtt,Piarulli:2019pfq} than used here to constrain our models. This is further amplified by the average mass of stars with a central density of $n_B=2\,\nsat$, the uppermost density to which \chiEFT\ might be reliable \cite{Tews:2018kmu,Drischler:2020yad}, indicated by the pink dots in Fig.~\ref{fig:chipt_sample_MR}. A star that light can therefore be completely described within \chiEFT\ and does not require a high-density extension using RMFTs.
In Ref.~\cite{superlight}, the authors present various mass-radius curves extracted from the same \chiEFT\ data shown in this paper that seem to meet all observational constraints at the $68\,\%$ credibility level. These curves are not obtained using an RMFT but rather by extending a simple parametrization of the \chiEFT\ band to higher densities \cite{Tews:2018kmu}. Within our framework, we were not able to fit an RMFT to the corresponding curves shown in Ref.~\cite{superlight}. 

\section{Hybrid Equations of State}\label{sec:hybrid}
In this section, we show that a soft, low-density nuclear EOS combined with a phase transition to a stiffer phase, e.g., quark matter, allows us to accommodate all astrophysical constraints at the $68\,\%$ credibility level. By augmenting the nuclear EOS with a low-density first-order phase transition to a quark matter EOS, we can circumvent a potential softening of the EOS due to hyperonic degrees of freedom because the quark matter phase sets in before a significant hyperon fraction builds up \cite{Bednarek:2011gd}. The mass-radius curve then contains a branch of hybrid stars with an inner core of quark matter and an outer core consisting of ordinary nuclear matter. 

To construct a hybrid EOS we use the constant speed of sound (CSS) model developed in Ref.~\cite{Alford:2013aca}. We use this model to attach an EOS with a constant speed of sound to a nuclear EOS in a thermodynamically consistent way, i.e.,~we demand that pressure and baryon chemical potential vary smoothly across the phase transition and that the baryon density is a monotonically increasing function with respect to the baryon chemical potential. The energy density as a function of the pressure (i.e.,~the EOS) is given by
\begin{equation}
\varepsilon(P) = \left\{\!
\begin{array}{ll}
\varepsilon_{\rm NM}(P) & P\leq\ptrans \\
\varepsilon_{\rm NM}(\ptrans)+\Delta\varepsilon+c_{\rm QM}^{-2} (P-\ptrans) & P\geq\ptrans
\end{array}
\right.\ ,
\label{eqn:EoSqm1}
\end{equation}
 where $\varepsilon_{\rm NM}(P)$ is the nuclear EOS. This EOS has three independent parameters: 1) the transition pressure $\ptrans$ where the quark matter phase becomes energetically preferred, 2) the jump in the energy density $\Delta\varepsilon$ which determines the strength of the first order phase transition, and 3) the constant speed of sound in the quark matter phase $c_{\rm QM}$, which determines the stiffness of the quark matter part of the EOS.The assumption of a (roughly) constant speed of sound is supported by perturbative QCD and NJL-model-based calculations presented in Refs.~\cite{Agrawal:2010er, Kurkela:2010yk,Bonanno:2011ch,Lastowiecki:2011hh,Zdunik:2012dj}. Although there are more sophisticated approaches to hybrid EOSs (see e.g.,~\cite{Ranea-Sandoval:2015ldr,Fraga:2022yls}), our simple approach allows us to easily examine the parameter space and show how hybrid models can accommodate the low mass and radius measurement of the compact object in HESS J1731-347. 
 
For the nuclear part of our hybrid EOSs, we choose a soft nucleonic RMFT that we call QMC-Soft, with a radius prediction  $R=11.47\,$km at $M=0.77\,\Msolar$. QMC-Soft's coupling constants and nuclear matter properties can be found in Tab.~\ref{tab:couplings}. For a detailed discussion of the couplings, the Lagrangian, and all relevant thermodynamic quantities, see Refs.~\cite{Alford:2022bpp,Alford:2023rgp}. We follow the same procedure for attaching a crustal EOS described in Sec.~\ref{sec:nuclear}.
\begin{table*}
\begin{tabular}{lcccccccccccc}
\hline
& $g_\si$ & $g_\om$ & $g_\rho$ & $b$ & $c$ & $b_1$ & $B$ & $n_\text{sat}$ & $\mathcal{E}(n_\text{sat})$ & $\kappa(n_\text{sat})$  & $J$ &$L$ \\[-0.3ex]
&     &     &     &     &     &   & [MeV$^4$] & [$\fm^{-3}$] & [MeV] & [MeV] & [MeV] & [MeV] \\ 
\hline
\textbf{QMC-Soft} & 6.58 & 6.56 & 10.87 & 0.0051 & 0.0949 & 13.357 & -933761 & 0.158 & -16.01 & 249 & 32.19 & 42.43  \\

\hline
\end{tabular}
\caption{
Couplings, pressure offset $B$, and selected properties for the QMC-Soft nuclear part of the hybrid EOSs. The exact definition of all quantities can be found in Ref.~\cite{Alford:2022bpp}.
}
\label{tab:couplings}
\end{table*}
The binding energy per nucleon for pure neutron matter and isospin-symmetric nuclear matter of QMC-Soft is shown in Fig.~\ref{fig:chipt_sample}. The properties of isospin-symmetric nuclear matter at saturation density, e.g., the symmetry energy $J$ and its slope $L$ at saturation density are within experimental constraints, e.g.,~Ref.~\cite{Drischler:2020hwi}. Because the pressure from QMC-Soft does not rise rapidly as density increases, this nuclear EOS can not support a two-solar-mass neutron star, as shown in Fig.~\ref{fig:chipt_sample_MR}.

We vary the three parameters of the CSS EOS to study the possible astrophysical predictions and test them against all astrophysical constraints described at the beginning of Sec.~\ref{sec:nuclear}. For simplicity, we translate the transition pressure $\ptrans$ to the corresponding transition baryon density $n_\mathrm{tr}$. The initial parameter range we consider is $n_\mathrm{tr}\in [2\nsat,3\nsat]$, $c^2_\mathrm{QM}\in[0.36,1]$, and $\Delta\varepsilon/\varepsilon_\mathrm{tr}\in[0.003,0.84]$. We further restrict the parameter space by discarding all models that do not predict a two-solar-mass star or form a detached or unstable or no hybrid branch. For a detailed mapping of the CSS parameter space, see Refs.~\cite{Alford:2013aca,Drischler:2020fvz}.
The lowest speed of sound squared in our study that can sustain a two-solar-mass compact star is $c_\mathrm{QM}^2\approx0.4$. Recent studies have shown that the speed of sound in heavy stars likely exceeds the conformal, high-density limit $c^2_\mathrm{conf}=1/3$, see, e.g., Refs.~\cite{Ecker:2022xxj,Altiparmak:2022bke}. The bottom right panel of Fig.~2 of Ref.~\cite{Ecker:2022xxj} indicates a nearly constant mean value for the speed of sound squared in the core of a two-solar-mass star around $c^2\approx 0.4-0.5$. For simplicity, we fix the speed of sound squared to a value close to the radial average of the mean of the distribution in Ref.~\cite{Ecker:2022xxj}, $c_\mathrm{QM}^2=0.48$ and explore the remaining parameter space.

\begin{figure}
    \centering
    \includegraphics[width=1\hsize]{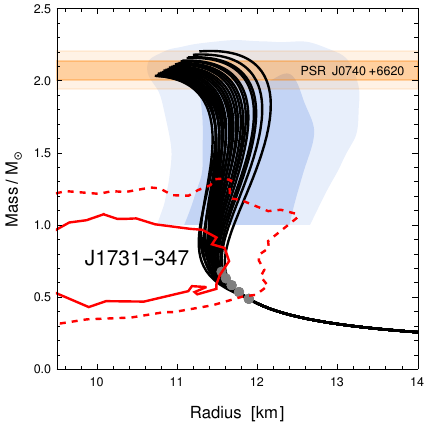}
    \caption{
      Mass-radius curves of hybrid stars with $c_\mathrm{QM}^2=0.48$ in the quark matter inner core and an outer core described by QMC-Soft, the nuclear model in Fig.~\ref{fig:chipt_sample}. The gray dots denote the transition point from the nuclear to the quark phase. All shaded contours are described in Fig.~\ref{fig:chipt_sample_MR}. This figure shows that hybrid models can easily fulfill all constraints on a stricter $68\,\%$ credibility level if paired with a soft nuclear model.}
    \label{fig:lowcshybrid}
\end{figure}

In Fig.~\ref{fig:lowcshybrid} we plot the mass-radius curves predicted by our hybrid EOSs with  $c_\mathrm{QM}^2=0.48$ that match \textbf{all} astrophysical constraints described in Sec.~\ref{sec:nuclear} at the $68\,\%$ credibility level. The transition densities (represented by gray dots) vary from $n_\mathrm{tr}=2\,\nsat$ to $n_\mathrm{tr}=2.4\,\nsat$, and the jump in the energy density varies from $\Delta\varepsilon/\varepsilon_\mathrm{tr}\approx0.004$ to $\Delta\varepsilon/\varepsilon_\mathrm{tr}\approx0.151$. We observe an inverse correlation between the transition density and the strength of the phase transition in the successful models: a higher transition density requires a smaller jump in the energy density (a weaker first-order phase transition) to obey all constraints; otherwise, detached branches will form.
We note that the speed of sound in quark matter, or even in nuclear matter, could be significantly higher, as shown in Ref.~\cite{Ecker:2022xxj}. It is also possible to construct weakly first-order hybrid EOSs with a speed of sound as low as $c_\mathrm{QM}^2\approx0.4$ that also obey all presented constraints at the $68\,\%$ credibility level. 

The exact model parameters for all EOSs presented in Fig.~\ref{fig:lowcshybrid} can be found online at \href{https://gitlab.com/ahaber}{gitlab.com/ahaber}. This includes the dimensionless tidal deformabilities of a $M=1.4\,\Msolar$ star for all the presented models in Fig.~\ref{fig:lowcshybrid}, which range from $\Lambda_{1.4\Msolar}=247$ to $\Lambda_{1.4\Msolar}=391$ and are well within the observational constraints from Refs.~\cite{LIGOScientific:2018cki,LIGOScientific:2018hze,Chatziioannou:2018vzf,Chatziioannou:2020pqz}. We compute the dimensionless tidal deformability via the second tidal love number \cite{Hinderer:2007mb}. There are corrections to the second tidal love number in the presence of strong first-order phase transitions that affect the I-Love-Q relations (see Fig.~1 within Ref.~\cite{Yagi:2013awa}) at the level of a few percent~\cite{Damour:2009vw,Paschalidis:2017qmb,Han:2018mtj,Carson:2019rjx,Takatsy:2020bnx}. Given that the hybrid models investigated in this work show a rather weak first-order phase transition and that the change in the dimensionless tidal deformability is well within known current constraints,  we neglect these corrections. We furthermore verify the consistency of these hybrid EOSs with constraints from perturbative QCD as derived in Refs.~\cite{Komoltsev:2021jzg,Gorda:2022jvk}, using the code the authors of Ref.~\cite{Gorda:2022jvk} provided publicly.

\section{conclusions}
\label{sec:conclusions}
We sample over 500 RMFTs constrained around nuclear saturation density by \chiEFT\ and inferred properties of isospin-symmetric nuclear matter to study how well they obey astrophysical constraints. We notice that the new observation of a very light CCO within supernova remnant HESS J1731-347 with mass \mbox{$M=0.77^{+0.20}_{-0.17}\,\Msolar$} and radius $R=10.4^{+0.86}_{-0.78}\,$km is barely compatible with our sampled RMFTs at the $95\,\%$ credibility level if other well-established observational constraints are taken into account. We are not able to construct an RMFT within the chosen model parameter space that obeys all constraints listed at the beginning of Sec.~\ref{sec:nuclear} at the $68\,\%$ credibility level. 

If future measurements confirm the mean and reduce the mass-radius uncertainty contour of the CCO, there will be tension between \chiEFT\ and astrophysical measurements. The core of neutron stars with $M\approx0.77\,\Msolar$ can be completely described by \chiEFT\, limiting any uncertainties introduced by using RMFTs. This might allow us, in the future, to use astrophysical observations to constrain the low energy constants of \chiEFT.

In the third section of this work, we present an alternative solution: a hybrid star with an outer core of nuclear matter and an inner core of quark matter. We show that such hybrid models can easily meet all constraints at the $68\,\%$ credibility level if the transition density from nuclear to quark matter takes place below $n_{\mathrm{tr}}\approx 2.5\,\nsat$ and does not require a strong first-order phase transition. To also support a heavy, two-solar-mass star, the speed of sound squared in quark matter must be above $c^2_\mathrm{QM}\approx0.4$. The speed of sound of our hybrid EOSs matches well with the bottom right panel of Fig.~2 within Ref.~\cite{Ecker:2022xxj}: a nearly constant speed of sound in the inner part of the star (which we model with a CSS quark matter EOS) and a rapid drop off in the outer regions (which we model with the nuclear QMC-Soft RMFT). 

Our results, together with other works like Ref.~\cite{Paschalidis:2017qmb}, emphasize that the existence of a quark matter core opens up the possibility of a softer nuclear EOS because the nuclear EOS does not have to support two-solar-mass compact stars. Neglecting the possibility of a quark matter core might therefore lead us to wrongly exclude soft nuclear models.

\bigskip
{\centering\bf Acknowledgements \par}
We thank Mark G.~Alford, Sophia Han, Ingo Tews, and Ziyuan Zhang for their input. 
This research was partly supported by the U.S. Department of Energy, Office of Science, Office of Nuclear Physics, under Award No.~\#DE-FG02-05ER41375.
\clearpage
\bibliography{superlight}

\end{document}